# Flexoelectricity-driven softening of bend elasticity leads to spontaneous chiral symmetry breaking in a polar fluid

Aitor Erkoreka [1], Josu Martinez-Perdiguero [1], Luka Cmok [2,3], Ema Hanžel [2,3], Jordan Hobbs [4], Calum J. Gibb [5], Richard J. Mandle [4,5], Nerea Sebastián [2], and Alenka Mertelj [2,*]

[1] *Department of Physics, Faculty of Science and Technology, University of the Basque Country EHU, Bilbao, Spain*
[2] *Jožef Stefan Institute, Ljubljana, Slovenia*
[3] *Faculty of Mathematics and Physics, University of Ljubljana, Ljubljana, Slovenia*
[4] *School of Physics and Astronomy, University of Leeds, Leeds, United Kingdom*
[5] *School of Chemistry, University of Leeds, Leeds, United Kingdom*

(Received 19 February 2026; accepted 17 September 2026; published 21 October 2026)

The origin of the recently observed spontaneous chiral symmetry breaking in polar fluids composed of achiral molecules is an unsolved problem, raising fundamental questions about how heliconical structures emerge in such systems. Here, we investigate the pretransitional fluctuations leading to the formation of the spontaneously chiral twist-bend ferroelectric nematic phase using dielectric spectroscopy, light scattering, and small-angle x-ray scattering. We observe simultaneous softening of the bend elastic constant and the emergence of a collective dielectric mode on approaching the transition. By developing a theoretical model, we show that these phenomena are signatures of a flexoelectricity-driven transition arising from the coupling between electric polarization and bend deformation.

## I. INTRODUCTION

In recent years, the discovery of the ferroelectric nematic ($N_F$) phase, a true three-dimensional ferroelectric fluid, has marked a major breakthrough, establishing a new perspective on the emergence of long-range polar order in soft condensed matter [1–5]. In this phase, the longitudinal dipole moments of achiral molecules spontaneously align roughly parallel to each other, thus creating a large electric polarization **P** (where $|\mathbf{P}| \sim 5\ \mu\mathrm{C/cm}^2$), which breaks the inversion symmetry of the director **n**, the unit vector that specifies the average direction of molecular orientation. Although often described as a uniaxial phase, it forms polydomain structures [4] which, under certain planar anchoring conditions, include domains with twisted polarization, driven by a larger reduction in electrostatic energy than the accompanying increase in elastic energy [6]. Following the groundbreaking discovery of the $N_F$ phase, additional related mesophases combining different degrees of orientational, positional, and polar order have been identified, making up the so-called ferroelectric nematic realm [7–11]. Particularly noteworthy are those systems exhibiting spontaneous chiral symmetry breaking despite being composed of achiral building blocks, like the ferroelectric twist-bend nematic ($N_{TBF}$, also called heliconical ferroelectric nematic or $^{HC}N_F$) and polar heliconical smectic C ($SmC_P^H$) phases [12–15]. Both phases are spontaneously polar and chiral in the bulk, where the helix pitch $p_h$ is on the order of the wavelength of visible light, the difference being that the latter additionally exhibits smectic (one-dimensional quasi-long-range positional) order. These discoveries raise a fundamental question of what drives the chiral symmetry breaking in ferroelectric nematic liquids made of achiral constituents.

The spontaneous chiral symmetry breaking in systems made of achiral building blocks is also observed in some conventional nematic ($N$) liquid crystals made of flexible dimers that exhibit a phase transition to the twist-bend nematic phase, a modulated nematic phase with heliconical structure and nanoscale modulation period ($N_{TB}$) [16–18], also referred to as polar-twisted nematic phase ($N_{PT}$) [19,20]. The pretransitional behavior at the $N$-$N_{TB}$ transition is governed by a softening of the bend elastic constant $K_3$ [16–18]. Different models, both phenomenological and microscopic, have been developed over the years to explain this behavior [21–24]. The overall softening appears to be the result of steric effects related to the bent molecular shape of the mesogens and not due to emergent biaxial order through flexoelectric coupling, as some theories suggested [24,25]. This is in stark contrast to polar liquid crystals like ferroelectric nematic liquids, where the transition from the high-temperature $N$ phase to the intermediate antiferroelectric splay-modulated $N_S$ phase [3,26,27], characterized by simultaneous pretransitional softening of the splay elastic constant $K_1$ and divergence of electric susceptibility, is driven by the flexoelectric coupling between splay deformation and electric polarization [3]. It should be noted that, in the literature, the $N_S$ phase is also named M2 [2], $N_X$, $SmZ_A$ [28], $N_{AF}$, and $M_{AF}$ [13].

*Contact author: alenka.mertelj@ijs.si

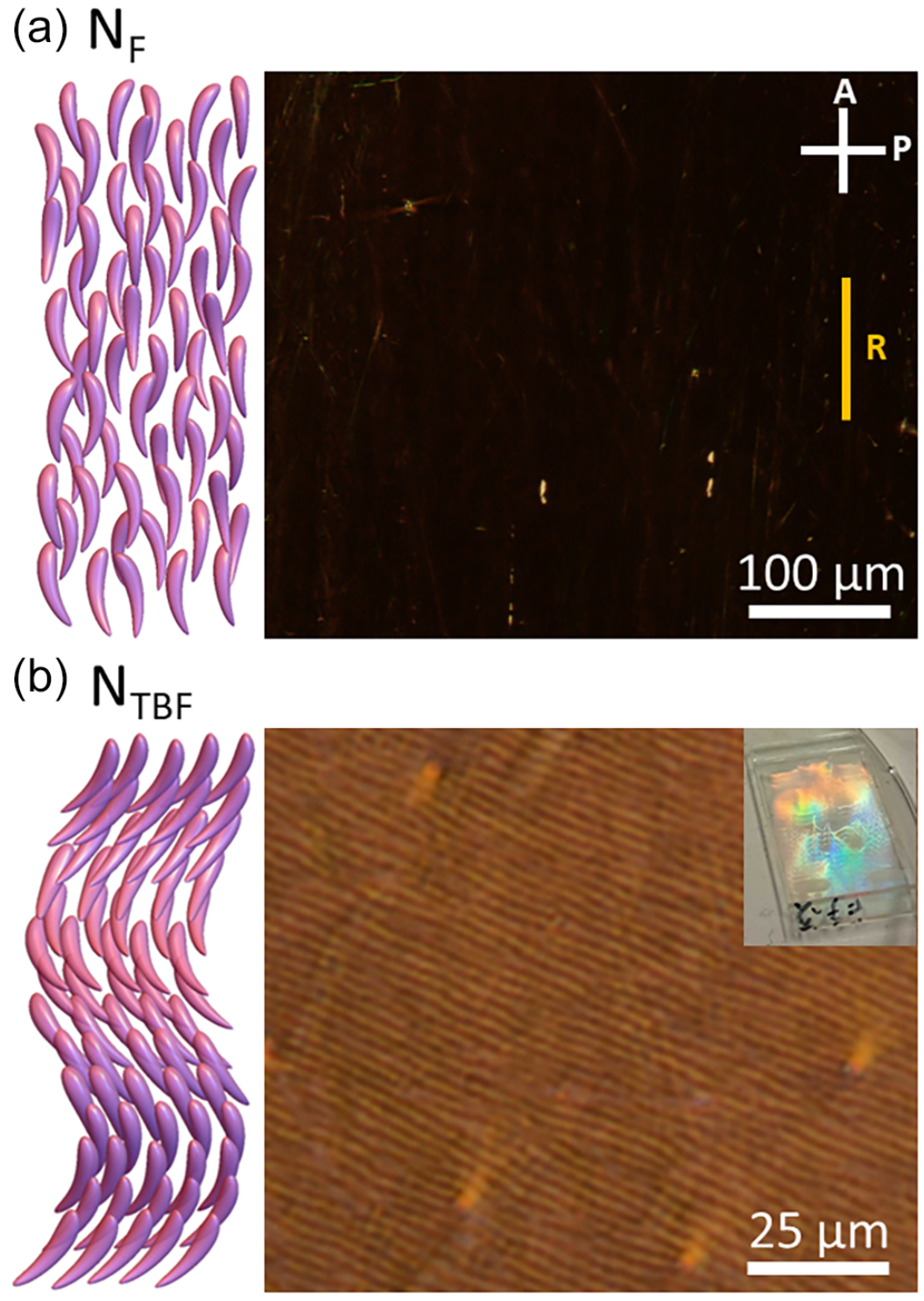

FIG. 1. Schemes and polarized microscope images of the material F7i in a 10-µm-thick cell with parallel rubbing. (a) $N_F$ at 32 °C and (b) $N_{TBF}$ just below the $N_F$-$N_{TBF}$ phase transition. P, A, and R denote the polarizer, analyzer, and rubbing directions, respectively. Inset (b) shows a photograph of the cell exhibiting strong Bragg scattering.

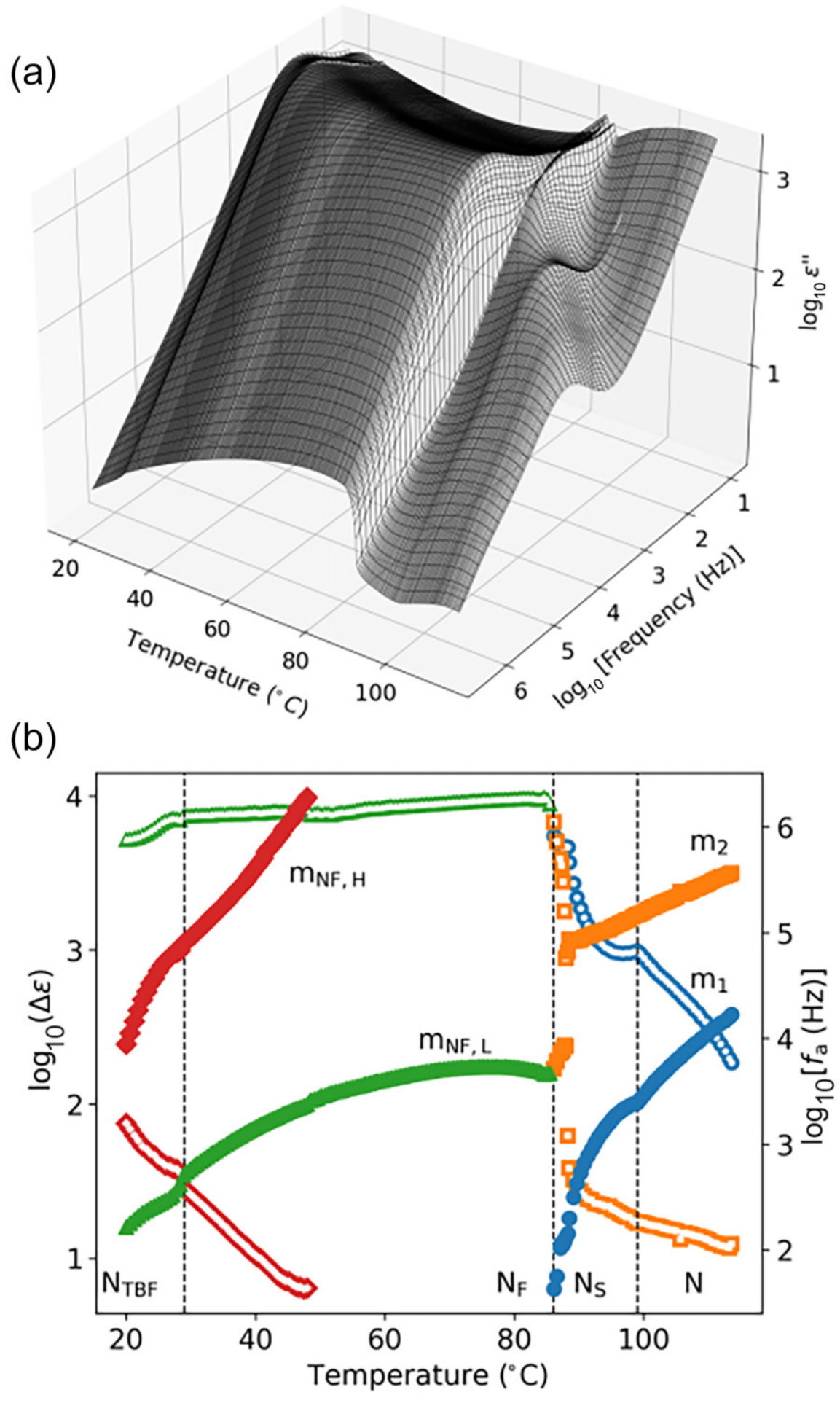

FIG. 2. (a) 3D plot of the dielectric absorption spectra vs temperature. (b) Temperature dependence of the dielectric strengths ($\Delta\varepsilon$, empty symbols) and frequencies of maximum absorption ($f_a$, full symbols) of the different relaxation processes.

In this paper, we propose that the increasing tendency toward polar biaxial order, which interacts with bend fluctuations through flexoelectric coupling, drives the chiral symmetry-breaking transition to the $N_{TBF}$ phase. In particular, we connect the emergence of biaxial order to the development of presmectic order.

## II. EXPERIMENTS

The studied material F7 (Appendix A, [29]) exhibits an $N_F$–$N_{TBF}$ phase transition at 29.5 °C. Figure 1 shows schematics of the uniform molecular arrangement in the $N_F$ and the heliconical structure of the $N_{TBF}$ phase, together with their corresponding textures as observed by polarized optical microscopy.

Firstly, in order to probe the dipolar fluctuations occurring along the entire phase sequence, we measured the complex dielectric permittivity $\varepsilon^*(f) = \varepsilon'(f) - i\varepsilon''(f)$ in the frequency range 10 Hz–3 MHz (Appendix B). The temperature evolution of the imaginary component of the dielectric spectra can be found in Fig. 2(a). To quantitatively analyze and interpret the dielectric relaxation processes, the spectra at each temperature were fitted to the Havriliak-Negami equation (Appendix B). The obtained strengths and frequencies of maximum absorption are shown in Fig. 2(b). The behavior in the $N$ and $N_S$ phases is similar to that previously found in other ferroelectric nematogens. Usually, these molecules tend to align homeotropically in untreated gold electrodes, as also seems to be the case here judging by the large amplitude of process $m_1$. This mode, corresponding to the rotation of individual molecules around their short axis, becomes collective in the $N_S$ phase and exhibits soft-mode behavior, thereby driving the transition to the lower-temperature $N_F$ phase. Process $m_2$, on the other hand, was observed in the $N$ phase for RM734 in Ref. [3], and more recently for a mixture in Ref. [30]. Its interpretation is currently unclear, although its higher frequency and Arrhenius-type temperature dependence, with an activation energy of ~74 kJ/mol (Appendix B), suggest it is related to more localized molecular motions. The transition to the $N_F$ phase is characterized by a strong dielectric response, where a collective relaxation process $m_{NF,L}$, i.e., the Goldstone mode, can be identified at low frequencies. It is well known that, in order to minimize the depolarization field, the phase transition is accompanied by a change of director orientation from homeotropic to planar [31–33]. There is an additional mode at lower frequencies, which was fitted in order to correctly deconvolute all the contributions to the dielectric spectra. Although this process could be associated to conductivity effects or Maxwell-Wagner polarization processes well known in dielectric spectroscopy [34], from the theoretical point of view an ionic mode coupled to the splay fluctuations is expected (Appendix F). It should be stressed that an undoped sample was used for the dielectric measurements, so these ions correspond to impurities in the system and their contribution may be small. As the temperature is lowered, however, another process $m_{\mathrm{NF,H}}$ enters the available frequency window and can be deconvoluted. Close to the transition to the $N_{\mathrm{TBF}}$ phase, it slows down considerably and its amplitude increases. We propose that it is related to the rotation of molecules around their long axis, which becomes collective due to the growth of the correlation length of the biaxial order, as will be discussed later.

Interested in the $N_{\mathrm{F}}$–$N_{\mathrm{TBF}}$ pretransitional behavior, we set out to study the evolution of orientational fluctuations by dynamic light scattering (DLS) and differential dynamic microscopy (DDM). In the apolar $N$ phase, the orientational fluctuations exhibit two branches: twist bend and splay bend. In the $N_{\mathrm{F}}$ phase, splay deformations increase the electrostatic self-energy and are therefore energetically disfavored, leading to suppression of the splay-bend mode (Appendix F). Consequently, in DLS and DDM experiments, the twist-bend branch prevails. The scattering geometries in the DLS experiments were chosen so that only pure bend or pure twist modes with a given scattering vector $q$ were measured. From the measurements, the temperature dependences of the scattered intensity $I_{\mathrm{T,B}} \propto (\Delta\varepsilon_{\mathrm{opt}})^2/(K_i q^2)$ and relaxation rates $1/\tau_i = K_i q^2/\eta_i$ were obtained. Here $i = 2, 3$ denote the twist and bend eigenmodes, respectively, $K_i$ and $\eta_i$ are the corresponding elastic constants and viscosities, and $\Delta\varepsilon_{\mathrm{opt}}$ the anisotropy of the optical dielectric tensor. The twist viscosity $\eta_2$ equals the rotational viscosity $\gamma_1$, while the bend viscosity is smaller due to backflow [35]. Figure 3 shows the temperature dependence of the bend and twist diffusivities $K_i/\eta_i$, elastic constants (the latter normalized at 55 °C, and the ratio $K_3/K_2$. The latter was determined by combining DDM and DLS results (see Appendixes C and D). It is clear that, while $K_2$ stays practically constant close to the phase transition temperature, $K_3$ exhibits a pronounced softening.

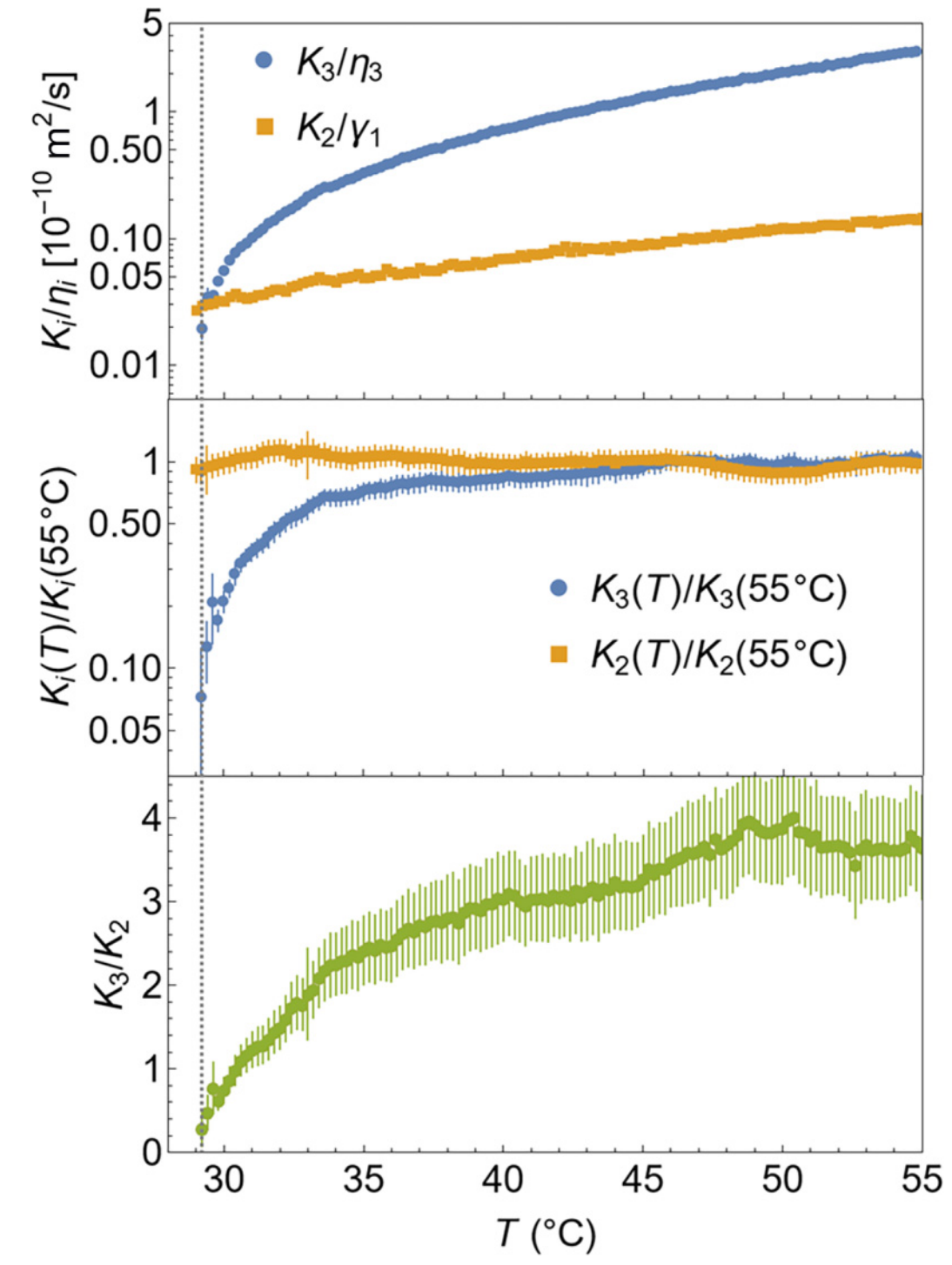

FIG. 3. Temperature dependence of the bend diffusivity $K_3/\eta_3$ and twist diffusivity $K_2/\gamma_1$ (top), normalized bend elastic constant $K_3$ and twist elastic constant $K_2$ (middle), and ratio $K_3/K_2$ (bottom) in the $N_{\mathrm{F}}$ phase. The dotted vertical line denotes the phase transition between the $N_{\mathrm{F}}$ and $N_{\mathrm{TBF}}$ phases.

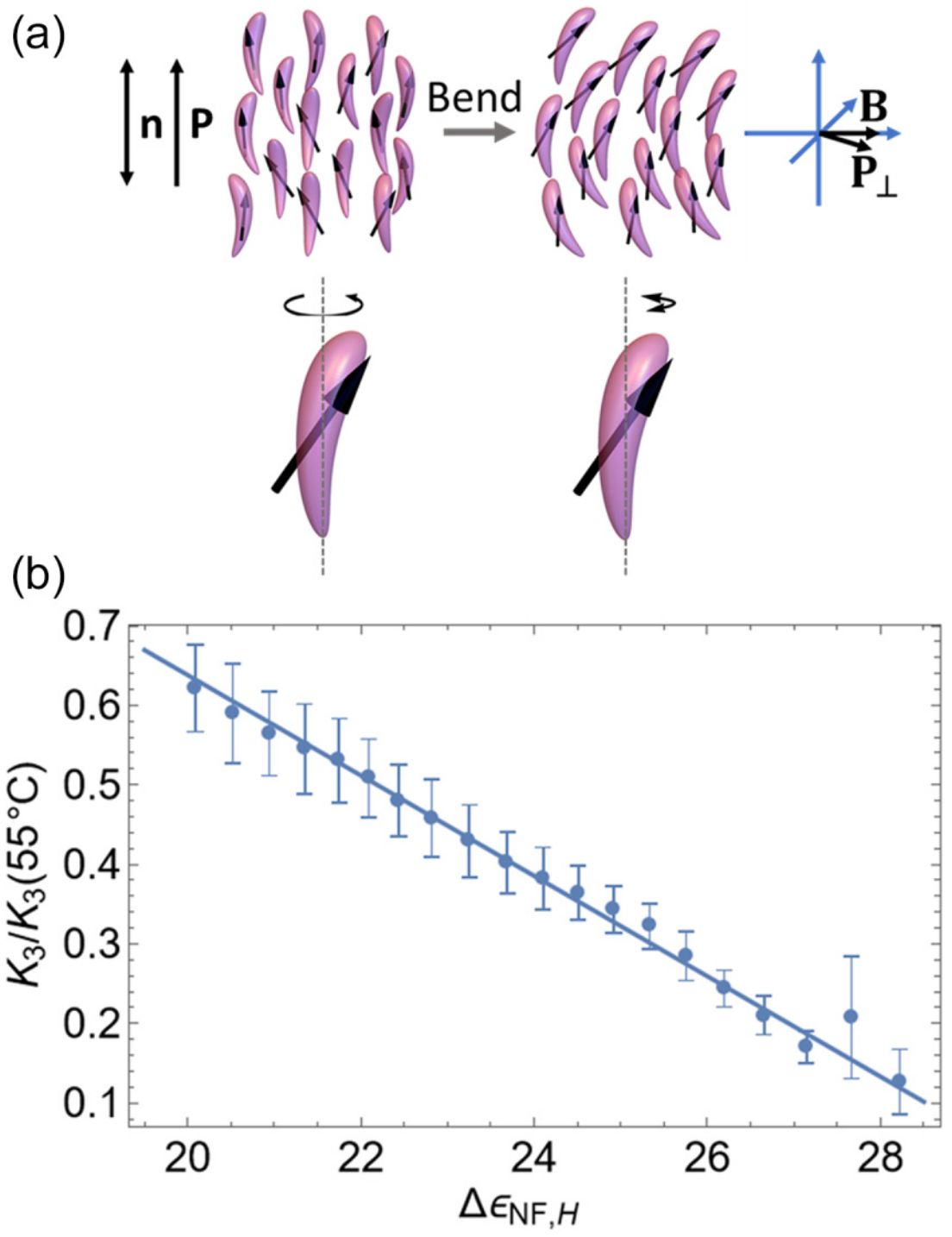

FIG. 4. (a) Schematics of the bend flexoelectric effect in the $N_{\mathrm{F}}$ phase. Here, **b** is the vector related to shape polarity (see text). (b) Dependence of $K_{3,\mathrm{eff}}/K_{3,\mathrm{eff}}$ (55 °C) on the amplitude of the mode $\Delta\varepsilon_{\mathrm{NF,H}}$. The line is a linear fit to the data yielding a value of $\gamma_B^2\varepsilon_0/K_{3,\mathrm{eff}}$ (55 °C) $= 0.063 \pm 0.002$.

## III. DISCUSSION

The simultaneous pretransitional softening of the bend elastic constant and emergent collective dielectric mode is a characteristic of a flexoelectricity-driven phase transition. It is a bend analogous to a ferroelectric-ferroelastic phase transition and can be described in a similar way. Assuming that, in equilibrium, the $N_{\mathrm{F}}$ phase is uniform and uniaxial, it can be described by the polarization vector $\mathbf{P} = P_0\mathbf{n}$, where $P_0$ is the polarization amplitude and the director $\mathbf{n}$ a unit vector. While the splay flexoelectric effect couples $\mathbf{P}$ to splay deformation, the bend flexoelectric effect couples the polarization perpendicular to $\mathbf{n}$, i.e., $\mathbf{P}_\perp$, with the bend deformation [36]. Its origin lies in the bent shape of the molecules. In the presence of bend deformation, such bent molecules pack more efficiently when they exhibit orientational order along their long axes [Fig. 4(a)] and, as a consequence, their rotational motion is hindered. Conversely, the presence of orientational order along the long axes promotes bend deformation. As the phase is uniaxial, then $\langle \mathbf{P}_\perp \rangle = 0$. Following Longa and Pająk [23], this polarization does not necessarily correspond to an electric polarization but may instead be related to shape polarity, i.e., polar biaxial order **b**, defined as a vector order parameter describing the orientational order of the (banana-shaped) molecules around their long axis [Fig. 4(a)]. If such a long-range order exists, the phase is biaxial. Furthermore, if these molecules possess a component of dipole moment perpendicular to their long axes, such an order will also result in a nonzero $\mathbf{P}_\perp$. While in an apolar $N$ phase $\mathbf{P}_\perp$ is parallel to **b**, in the $N_\mathrm{F}$ phase this is no longer the case [Fig. 4(a)]. In general, there is a finite angle between both vectors that depends on the angle between the molecular bend and electric dipole; however, the magnitudes are proportional, $|\mathbf{P}_\perp| \propto |\mathbf{b}|$. Here, we are interested in bend fluctuations coupled to fluctuations of $\mathbf{P}_\perp$, that is, small periodic fluctuations of $\mathbf{P}_\perp$ and the angle $\varphi$, which describes the director $\mathbf{n} = [\sin(\varphi(\mathbf{q},t)), 0, \cos(\varphi(\mathbf{q},t))]$, with $\mathbf{q} = (0,0,q)$ around the uniform state with $\mathbf{P}_\perp = 0$ and $\mathbf{n} = (0,0,1)$. The relevant part of the free energy density describing these fluctuations is

$$f = \tfrac{1}{2}K_3[\mathbf{n}\times(\nabla\times\mathbf{n})]^2 - \gamma_B\mathbf{n}\times(\nabla\times\mathbf{n})\cdot\mathbf{P}_\perp + \tfrac{1}{2}a_\perp P_\perp^2 + \tfrac{1}{2}K_\perp(\nabla P_\perp)^2. \quad (1)$$

The first term is the bend elastic energy density, the second is the flexoelectric term with $\gamma_B$ being the bend flexoelectric coefficient, and the last two terms are the lowest-order terms in $\mathbf{P}_\perp$ and its gradients allowed by symmetry. This leads to coupled director and $\mathbf{P}_\perp$ fluctuation modes with the relaxation rates [37]

$$\frac{1}{\tau_{01}} = \frac{K_3 - \frac{\gamma_B^2}{a_\perp}}{\eta_3}q^2, \quad (2)$$

$$\frac{1}{\tau_{02}} = \frac{a_\perp}{\eta_P} + \left(\frac{K_\perp}{\eta_P} + \frac{\gamma_B^2}{\eta_3 a_\perp}\right)q^2. \quad (3)$$

The first is a hydrodynamic mode and is the one observed in light-scattering experiments as a bend mode. The flexoelectric coupling causes the bend elastic constant to be replaced by an effective one $K_{3,\mathrm{eff}} = K_3 - \gamma_B^2/a_\perp$. The second mode is dominantly a collective polarization mode measured (at $q = 0$) in dielectric spectroscopy. Analogously to the splay case [3,37], the amplitude of this dielectric mode is related to $a_\perp = \frac{1}{\varepsilon_0\Delta\varepsilon}$ and consequently $K_{3,\mathrm{eff}} = K_3 - \gamma_B^2\varepsilon_0\Delta\varepsilon$. In normal nematic liquid crystals, the mode associated with the rotation of molecules around their long axis is noncollective, fast, and has a small amplitude, so the coefficient $a_\perp$ is large and biaxial order is not favorable. However, if such a mode becomes collective, it slows down and its amplitude starts to grow, and causes $K_{3,\mathrm{eff}}$ to decrease. We propose that the $m_{\mathrm{NF,H}}$ mode in Fig. 2(b) is causing the softening of the bend elastic constant via bend flexoelectric coupling. Figure 4(b) shows the dependence of the measured $K_{3,\mathrm{eff}}/K_{3,\mathrm{eff}}$ (55 °C) on the amplitude of the mode $\Delta\varepsilon_{\mathrm{NF,H}}$ which is linear as predicted by the theory. $K_{3,\mathrm{eff}}$ approaches zero and the system becomes unstable toward bend deformation. This happens when $a_\perp = \gamma_B^2/K_3$ is still positive, i.e., before biaxial order on its own becomes energetically favorable. As discussed by Dozov [38], because it is not possible to fill space with homogeneous bend, this in the ordinary $N$ phase leads either to the twist-bend or, if $K_1 < 2K_2$, to the splay-bend phase. In the ferroelectric case, the splay deformation is associated with large electrostatic energy (Appendix F), so the heliconical structure of the $N_\mathrm{TBF}$ phase prevails in any case.

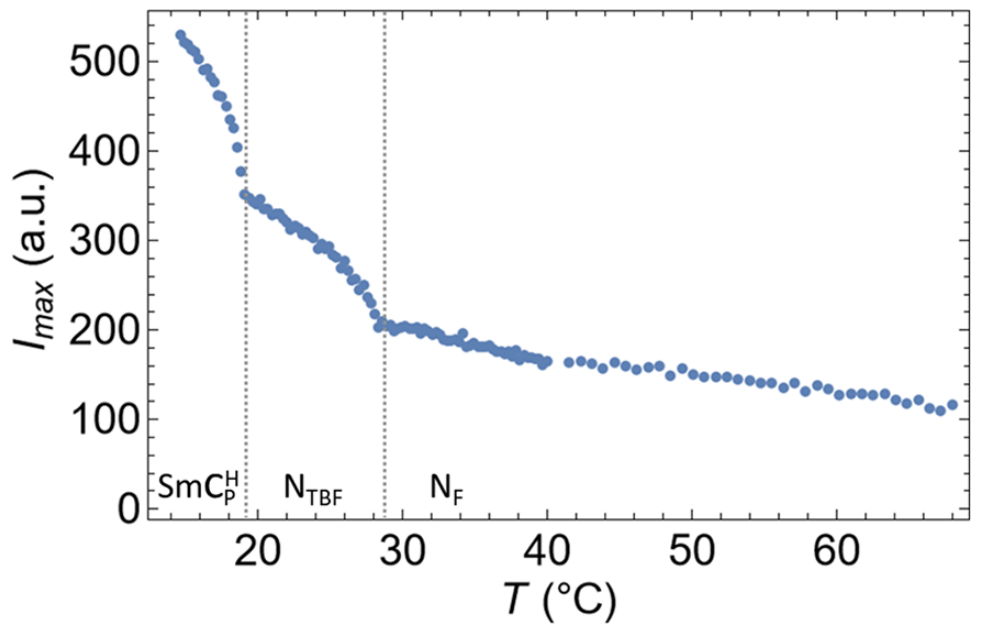

FIG. 5. Temperature dependence of the amplitude of the x-ray scattering peak that corresponds to local smectic order.

The coefficient $\gamma_B$ can be estimated from the linear fit in Fig. 4, which gives $\gamma_B^2\varepsilon_0/K_{3,\mathrm{eff}}$ (55 °C) $= 0.063 \pm 0.002$. Assuming $K_{3,\mathrm{eff}}$(55 °C) $\approx$ 10–50 pN, gives the flexoelectric coefficient $|e_3| = \varepsilon_0|\gamma_B| \approx$ 2–5 pC/m, which is a typical value of this coefficient in the conventional nematic phase [39]. The bend flexoelectric effect originates from the bent shape of the constituent molecules and is a common feature of the nematic phase. A slightly bent molecular conformation, even if subtle in some cases, is also present in compounds exhibiting the $N_\mathrm{TBF}$ phase, including the present material (see Appendix A) as well as other reported systems. For instance, in MUT_JK103 [12] it comes from torsional offsets between adjacent aromatic rings across the ester linkage, whereas in $n$BOE ($n = 4$–$6$) [13,40], despite the linear geometry of the alkyne spacer, weak steric and electronic interactions between adjacent aromatic units, together with the tilt of the terminal alkyl chain, result in a slightly bent overall conformation.

Upon cooling, the $N_\mathrm{TBF}$ phase undergoes a transition to $\mathrm{SmC_P^H}$. The small-angle x-ray scattering (SAXS) shows pretransitional smectic fluctuations already throughout the $N_\mathrm{F}$ phase [29]. Such pretransitional smectic order has also been identified previously in other compounds exhibiting the $N_\mathrm{TBF}$ phase [12,13,41] and appears to be a common feature of these materials. The amplitude of the peak that corresponds to local smectic order continuously grows with decreasing temperature and exhibits a kink at the $N_\mathrm{F}$–$N_\mathrm{TBF}$ transition (Fig. 5). The SmC phase is biaxial, so the SmC fluctuations are also fluctuations in biaxial order. These collective fluctuations are detected by broadband dielectric spectroscopy (BDS) because they carry a net dipole moment in the direction perpendicular to the director and they are coupled to bend through the flexoelectric effect. This explains the continued evolution of this mode in the $N_\mathrm{TBF}$ phase [Fig. 2(b)].

## IV. CONCLUSIONS

In summary, we have attempted to answer the question regarding the origin of spontaneous chiral symmetry breaking in polar fluids composed of achiral molecules. To this end, we conducted a series of dielectric, light scattering, and x-ray diffraction experiments to examine the pretransitional fluctuations leading to the appearance of a spontaneously chiral polar mesophase. We have simultaneously observed a pretransitional softening of the bend elastic constant and the emergence of a collective dielectric mode, and, by developing an appropriate theoretical model, we interpret this as a flexoelectricity-driven transition. The data are consistent with this interpretation, and the model shows that the increasing tendency toward polar biaxial order, when combined with even a slight bend in the shape of the constituent highly polar molecules, is sufficient to drive the transition.

## ACKNOWLEDGMENTS

A.E. and J.M.-P. acknowledge funding from the Basque Government Project No. IT1979-26 and from Project No. PID2023-150255NB-I00 from MCIU/AEI/10.13039/5011000-11033/FEDER, UE. J.H., C.J.G., and R.J.M. acknowledge funding from UKRI via a Future Leaders Fellowship, Grant No. MR/W006391/1. L.C., E.H., N.S., and A.M. acknowledge the support of the Slovenian Research and Innovation Agency, Grants No. P1-0192, No. J1-50004, and No. BI-VB/25-27-011.

## DATA AVAILABILITY

The data that support the findings of this article are openly available [42].

## APPENDIX A: MATERIALS

In the experiments, a binary mixture F7 composed of 70 mol % DIO and 30 mol % compound 1, from Ref. [14], with phase sequence $\mathrm{SmC}_\mathrm{P}^\mathrm{H}$ (20.9 °C), $N_\mathrm{TBF}$ (29.5 °C), $N_\mathrm{F}$ (88.8 °C), $N_\mathrm{S}$ (99.4 °C) $N$ was used [29]. For the DLS and DDM experiments, 0.06 wt % of an ionic dye (modified rhodamine B) was added to the mixture F7, yielding the sample F7i, to achieve stable, uniform in-plane alignment of the director [Fig. 1(a)]. The addition of the dye did not affect the phase transition temperatures. The molecular structures of the employed compounds can be found in Fig. 6.

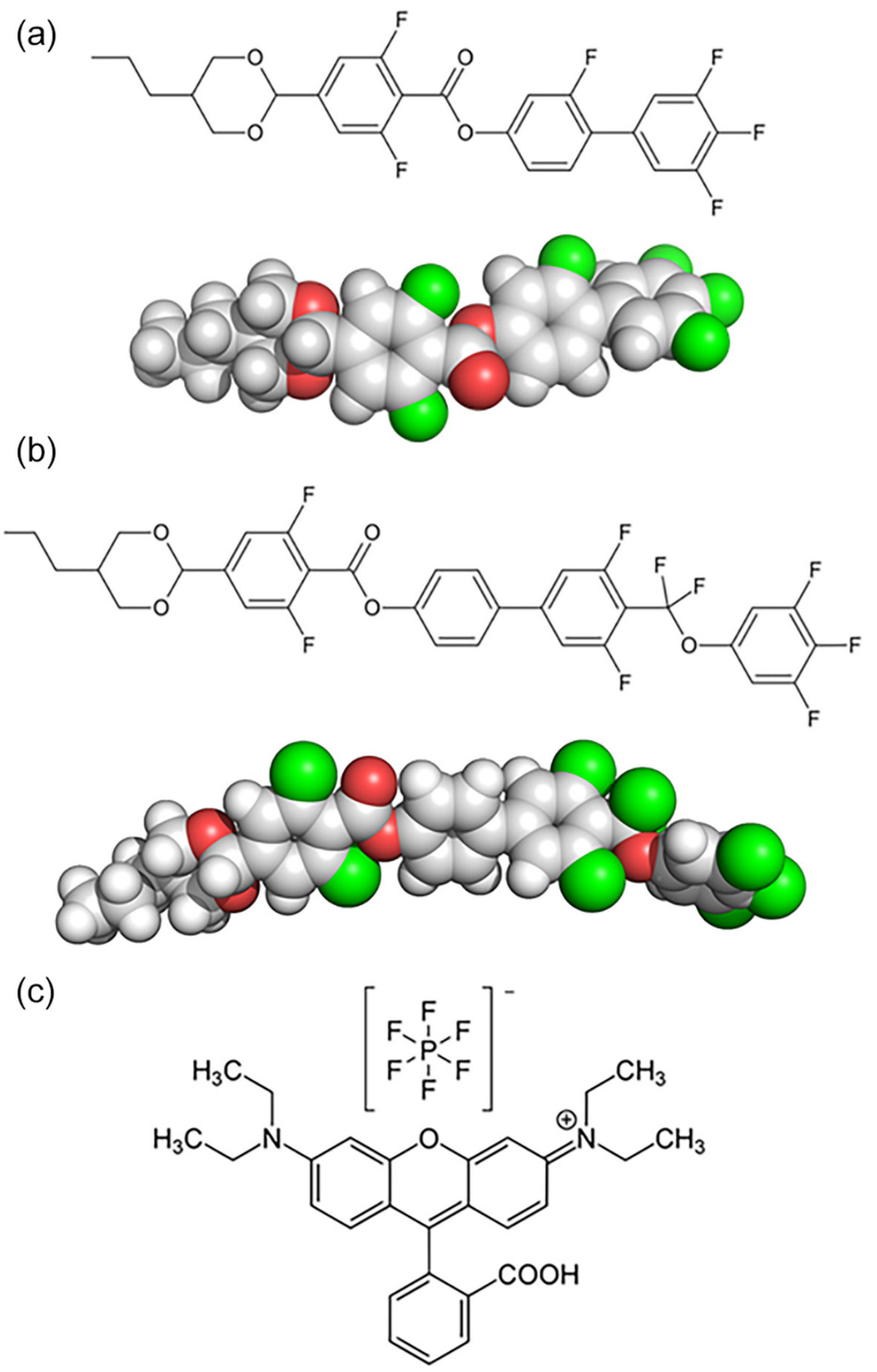

FIG. 6. Molecular structures of (a) DIO, (b) compound 1 (with its optimized geometry calculated at the B3LYP-GD3BJ/aug-cc-pVTZ level of density functional theory as detailed in Ref. [14] evidencing its slightly bent shape), and (c) modified RhB.

## APPENDIX B: BDS

The dynamic dielectric function $\varepsilon^*(f) = \varepsilon'(f) - i\varepsilon''(f)$ of F7 was measured in the spectral range $f = 10$ Hz–3 MHz using an Alpha-A impedance analyzer from Novocontrol Technologies GmbH. The sample was placed between two circular gold-coated brass electrodes 5 mm in diameter forming a parallel-plate capacitor, the separation being fixed at $d = 20$ μm by spherical silica spacers. The high conductivity of gold ensured reliable measurements up to the highest frequencies. This cell was then placed at the end of a modified HP 16091A coaxial test fixture, using a Quatro cryostat for temperature control. No polymer aligning layers were used on the electrodes because, although they can be useful to obtain a proper alignment of liquid-crystal molecules, they act as an additional large capacitance in series with the measurement cell and can lead to undesired charge accumulation, among other effects. The oscillator voltage was set to 0.03 $V_\mathrm{rms}$ to ensure a linear regime. Experiments were done on cooling from 115 °C at 0.25 K/min. The stray capacitance of the measurement circuit was subtracted from the measured capacitance, and the complex dielectric permittivity was obtained by dividing this value by the capacitance of the empty cell. Finally, the experimental curves were fitted to Havriliak-Negami (HN) relaxations with a conductivity term:

$$\varepsilon^*(f) = \sum_k \frac{\Delta\varepsilon_k}{\left[1 + \left(i\frac{f}{f_k}\right)^{\alpha_k}\right]^{\beta_k}} + \varepsilon_\infty + \frac{\sigma}{\varepsilon_0 (i2\pi f)^\lambda}, \qquad \text{(B1)}$$

where $\Delta\varepsilon_k$, $f_k$, $\alpha_k$, and $\beta_k$ are, respectively, the dielectric strength, relaxation frequency, and broadness exponents of mode $k$, $\varepsilon_\infty$ is the high-frequency dielectric permittivity, $\sigma$ is a measure of the conductivity, and $0 < \lambda \leqslant 1$. Fit examples can be found in Fig. 7.

## APPENDIX C: DLS

For the DLS experiments, the sample F7i was used to achieve uniform in-plane alignment of the director along the rubbing direction in a liquid-crystal cell (EHC, thickness 10 μm, parallel rubbed) in the $N_\mathrm{F}$ phase. The heating stage (Instec, HCS412W) was used to control the sample tempera-ture. The sample was cooled from 55 °C to room temperature at 0.02 K/min. In the DLS setup, a frequency-doubled diode-pumped Nd:YAG laser (532 nm, 80 mW attenuated for 100×), a single mode optical fiber with a GRIN lens connected to an ALV APD based "pseudo" cross-correlation detector, and an ALV-6010/160 correlator was used to obtain the autocorrelation function of the scattered light intensity. The direction and the polarization of the incoming and detected light were chosen so that pure twist and bend modes were observed [37]. The duration of a single measurement was 600 s. The measured intensity autocorrelation function $g_2$ was fitted by $g_2 = 2(1 - j_D)j_D g_1 + j_D^2 g_1^2 + y_0$, where the fitting parameters $j_D$ and $y_0$ are the ratio between dynamic and total scattered intensity, and background, respectively, while $g_1$ was either a single $g_1 = e^{-t/\tau_T}$ (for the twist geometry), or a double exponential function $g_1 = (1 - a_2)e^{-t/\tau_1} + a_2 e^{-t/\tau_B}$ (for the bend geometry). In the bend geometry, besides the dominating bend mode, a weaker and slower mode was observed of which relaxation rate $1/\tau_1$ did not show significant temperature dependence. In the temperature region below 35 °C, it became indistinguishable due to the increasing amplitude and slowing down of the bend mode. From the relaxation rates, the twist $K_2/\gamma_1$ and bend $K_3/\eta_3$ diffusivities were calculated. The scattered intensity $I_{2,3}$ of the twist and bend modes was determined as products $j_D I_{tot}$ and $a_2 j_D I_{tot}$, respectively, where $I_{tot}$ was the total detected intensity. As in the temperature range between 29 °C and 55 °C the refractive indices and, consequently, the anisotropy of the optical dielectric tensor are almost constant, $K_{2,3}(T)/K_{2,3}\,(55\,°C) \approx I_{2,3}\,(55\,°C)/I_{2,3}(T)$.

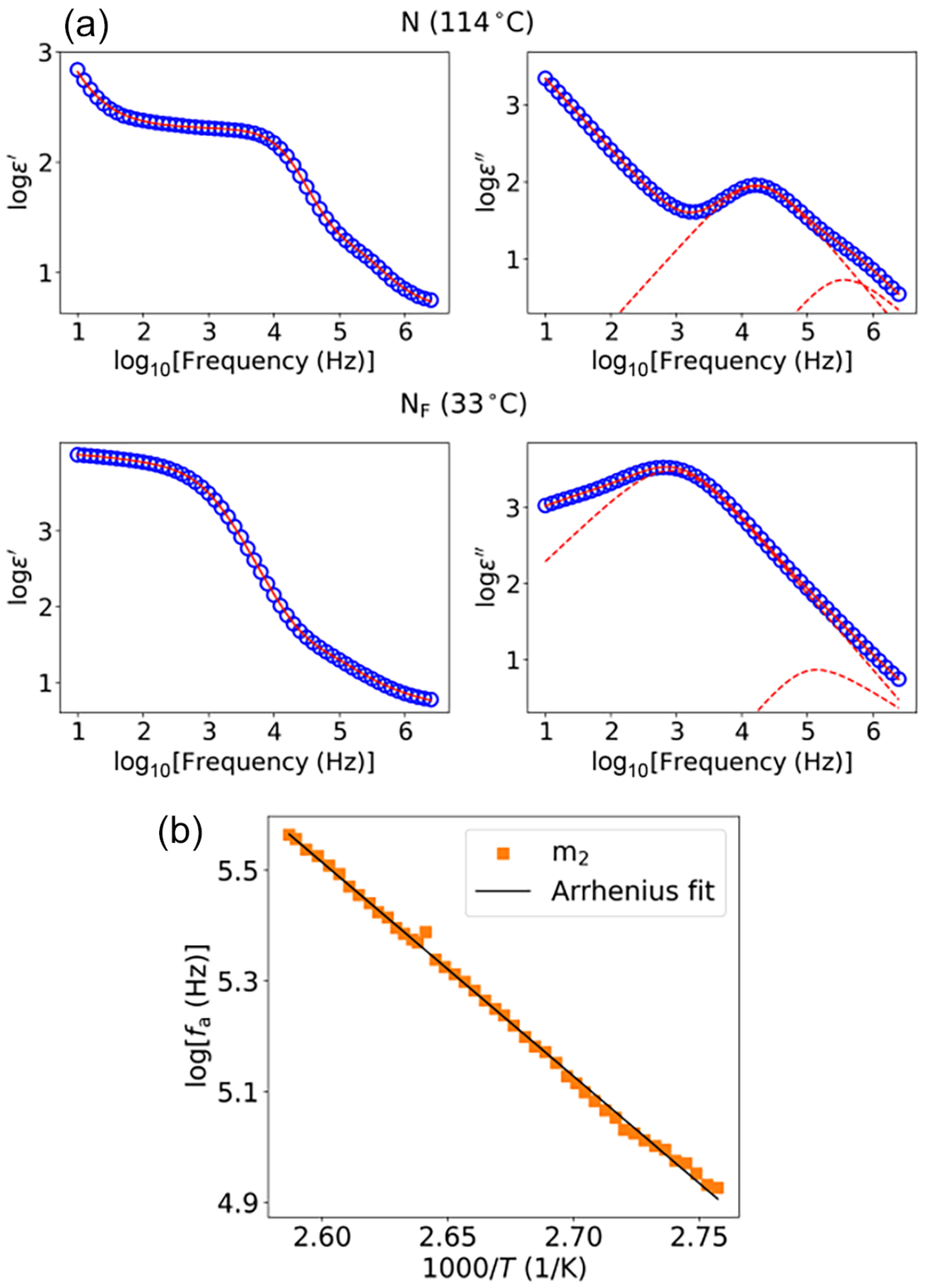

FIG. 7. (a) Examples of Havriliak-Negami fits at representative temperatures. (b) Arrhenius plot for process $m_2$ giving an activation energy $E_a \sim 74$ kJ/mol.

## APPENDIX D: DDM

In the DDM experiments, the same sample was used as in the DLS experiments. The heating stage (Instec, HCS302XY) was used to control the sample temperature. The sample was cooled from 35 °C to room temperature at 0.02 K/min. A cross DDM setup was used as described in [37]. The angles between the director $\mathbf{n}$ and the polarizer and analyzer were 80° and 0°, respectively. To calculate the normalized image-cross-correlation function an open-source package CDDM was used [43]. The correlation functions at all 2D scattering vectors $\mathbf{q}_{2D} = (q_\perp, q_\parallel)$ with enough signal showed one relaxation which was fitted by a single exponential decay, $g_{DDM} = a_0 e^{-t/\tau} + y_0$. In this geometry, the scattering from twist-bend mode prevails so the 2D dispersion curve $1/\tau(\mathbf{q}_{2D})$ can be attributed to this mode. Taking into account the relation between the scattering vector $\mathbf{q}$ and $\mathbf{q}_{2D}$, $\mathbf{q} = (q_\perp, q_\parallel, q_z)$ with $q_z \approx q_0 - (q_\perp^2 + q_\parallel^2)/(2k_0 n_e)$, its relaxation rate as a function of $\mathbf{q}_{2D}$ can be expressed in terms of five parameters:

$$\frac{1}{\tau_{TB}} = D_0 \times \frac{1 + a_{x2}q_\perp^2 + \frac{a_{x2}^2 q_\perp^4}{4} + a_{y2}q_\parallel^2 + \frac{1}{2}a_{x2}a_{y2}q_\perp^2 q_\parallel^2 + a_{y4}q_\parallel^4}{1 + \frac{a_{x2}q_\perp^2}{2} + b_{y2}q_\parallel^2}. \tag{D1}$$

The parameters depend on the optic and viscoelastic properties of the material: $D_0 = K_2 q_0^2/\gamma_1$, $a_{x2} = (2n_o)/(n_e q_0^2)$, $a_{y2} = [2n_o + n_e(-2 + k_{32} + \upsilon_{ca})]/(n_e q_0^2)$, $a_{y4} = \{[(-1 + k_{32})n_e + n_o][n_o + n_e(-1 + \upsilon_{ca})]\}/(n_e^2 q_0^4)$, and $b_{y2} = [n_o + n_e(-1 - \alpha_N^2 + \upsilon_{ca})]/(n_e q_0^2)$, with $n_o$ and $n_e$ being ordinary and extraordinary indices of refraction, respectively, $q_0 \approx (n_e - n_o)\frac{2\pi}{\lambda}$, $k_{32} = \frac{K_3}{K_2}$, $\upsilon_{ca} = \frac{\eta_a}{\eta_c}$, $\alpha_N^2 = \frac{\alpha_2^2}{\gamma_1 \eta_a}$, $\eta_{a,c}$ Miesowicz viscosities [35], $\alpha_2$ Leslie viscosity coefficient [35], and $\lambda$ wavelength of the LED diode used for the illumination (565 nm). All five fitting parameters can be reliably obtained from the fitting of $1/\tau(\mathbf{q}_{2D})$ only in the temperature region of 3 K above the phase transition. From the parameters, $k_{32}$ was extracted and compared to the $k_{32}(T)/k_{32}$ (55 °C) obtained from the DLS measurements [Fig. 3(b)] to determine $k_{32}$ (55 °C) $= 3.6 \pm 0.3$.

## APPENDIX E: SAXS

SAXS patterns of F7 were recorded using a Stoe Stadivari goniometer equipped with a Genix3D microfocus generator (Xenocs) and a Dectris Pilatus 100K detector. Monochromatic Cu $K\alpha$ radiation ($\lambda = 1.5406$ Å) was employed, setting the exposure time to 1 min. The temperature was varied using a nitrogen-gas cryostream controller (Oxford Cryosystems) achieving a temperature control within 0.1 °C. The material was introduced into a Lindemann capillary 0.6 mm in diameter. Each diffractogram was measured while rotating the capillary by 360° to avoid alignment effects. The peak amplitude shown in Fig. 5 is the peak height extracted from the difference between the SAXS diffractogram $I(T)$ and the SAXS diffractogram $I_N$ measured in the $N$ phase at 110 °C; examples are shown in Fig. 8.

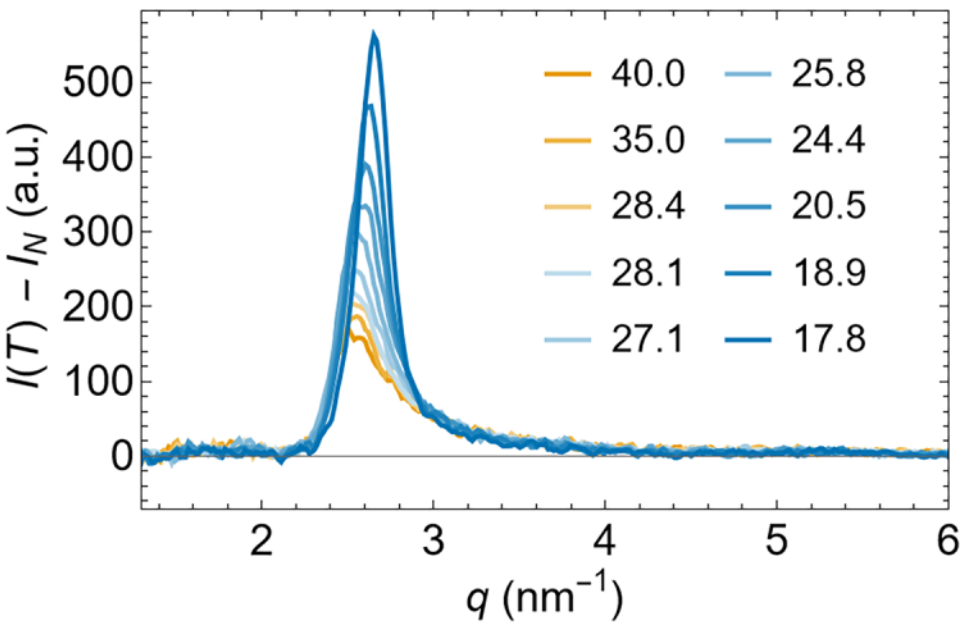

FIG. 8. Temperature evolution of the difference between SAXS intensity $I(T)$ and SAXS intensity $I_N$ measured in the $N$ phase at 110 °C showing growing smectic correlations. The labels are the corresponding temperatures in °C.

## APPENDIX F: ORIENTATIONAL FLUCTUATION MODES IN THE $N_F$ PHASE

As in the conventional nematic ($N$) phase, the orientational-fluctuation eigenmodes in the $N_F$ phase are also overdamped plane waves and have two branches: splay bend and twist bend. However, whereas in the $N$ phase the splay-bend and twist-bend modes at a given wave vector $\mathbf{q}$ have comparable relaxation rates and amplitudes [35], in the $N_F$ phase they become more distinct because they are affected differently by the electrostatic self-energy. The sources of this energy are bound charges given by $-\nabla \cdot \mathbf{P}$. In the $N_F$ phase, the polarization $\mathbf{P}$ is parallel to $\mathbf{n}$, $\mathbf{P} = P_0\mathbf{n}$, which means that the director splay deformation characterized by a nonzero $\nabla \cdot \mathbf{n}$ will increase the electrostatic energy of the system. In contrast, the twist and bend deformations, given by a nonzero $\mathbf{n} \cdot (\nabla \times \mathbf{n})$ and $\mathbf{n} \times (\nabla \times \mathbf{n})$, respectively, are electrostatically neutral. Consequently, the twist-bend fluctuation branch remains the same as in the $N$ phase, while the splay-bend branch is strongly affected. The increase of the free energy due to small periodic fluctuations around the equilibrium structure with the uniform director $\mathbf{n}_0 = (0, 0, 1)$ and polarization magnitude $P_0$ consists of several terms:

$$F = \int (f_{\rm el} + f_P + f_{\rm c} + f_{\rm ES})dV. \tag{F1}$$

The first term is the Frank-Oseen elastic energy,

$$f_{\rm el} = \tfrac{1}{2}K_1(\nabla \cdot \mathbf{n})^2 + \tfrac{1}{2}K_2[\mathbf{n} \cdot (\nabla \times \mathbf{n})]^2 + \tfrac{1}{2}K_3[\mathbf{n} \times (\nabla \times \mathbf{n})]^2. \tag{F2}$$

Here, $K_i$ are the splay ($i = 1$), twist ($i = 2$), and bend ($i = 3$) orientational elastic constants. The second term corresponds to the small fluctuations of the magnitude of $\mathbf{P}$, $\delta P = P - P_0$, and fluctuations of polarization perpendicular to $\mathbf{n}$, $\mathbf{P}_\perp$, and consists of the lowest-order by symmetry allowed terms in $\delta P$ and $P_\perp = |\mathbf{P}_\perp|$, and their gradients,

$$f_P = \tfrac{1}{2}a_\parallel(\delta P)^2 + \tfrac{1}{2}K_\parallel(\nabla \delta P)^2 + \tfrac{1}{2}a_\perp(P_\perp)^2 + \tfrac{1}{2}K_\perp(\nabla P_\perp)^2. \tag{F3}$$

The flexoelectric term couples the director fluctuations to the polarization fluctuations,

$$f_{\rm c} = -\gamma_{\rm S}\delta P(\nabla \cdot \mathbf{n}) - \gamma_{\rm B}\mathbf{n} \times (\nabla \times \mathbf{n}) \cdot \mathbf{P}_\perp, \tag{F4}$$

with $\gamma_{\rm S}$ and $\gamma_{\rm B}$ being the splay and bend flexoelectric coefficient, respectively.

The last term is the electrostatic self-interaction [44],

$$f_{\rm ES} = \tfrac{1}{2}\varepsilon_0 \nabla\Phi \cdot \nabla\Phi + \mathbf{P} \cdot \nabla\Phi - \rho_{\rm f}\Phi, \tag{F5}$$

where the electrostatic potential $\Phi$ is given by the Poisson equation, $\varepsilon_0\nabla^2\Phi = \nabla \cdot \mathbf{P} - \rho_{\rm f}$, and the electric field $\mathbf{E} = -\nabla\Phi$. Here, $\mathbf{P} = (P_0 + \delta P)\mathbf{n} + \mathbf{P}_\perp$ and $\rho_{\rm f}$ is the free ion density. In the cases $\delta P$ or/and $\mathbf{P}_\perp$ are not directly coupled to the dynamics of $\mathbf{n}$ and their response to the electric field is much faster than the dynamics of $\mathbf{n}$, $\delta P = E_\parallel/a_\parallel = (\varepsilon_\parallel - 1)\varepsilon_0 E_\parallel$ or/and $P_\perp = E_\perp/a_\perp = (\varepsilon_\perp - 1)\varepsilon_0 E_\perp$. The dielectric tensor is then $\boldsymbol{\varepsilon} = \varepsilon_\perp \boldsymbol{I} + (\varepsilon_\parallel - \varepsilon_\perp)\mathbf{n} \otimes \mathbf{n}$ with $\varepsilon_\parallel$ and $\varepsilon_\perp$ being its components along and perpendicular to $\mathbf{n}$, respectively. $\varepsilon_0$ is the vacuum permittivity.

The pure bend modes are discussed in the main text. Here, we focus on the pure splay and twist modes. The pure splay modes are observed when the wave vector is perpendicular to the equilibrium director $\mathbf{n}_0$, e.g., $\mathbf{q} = (q_\perp, 0, 0)$, and the director fluctuations lie in the plane spanned by $\mathbf{n}_0$ and $\mathbf{q}$. Here, we are interested in small periodic fluctuations of the angle $\varphi_{\rm s} = \varphi_{0\rm s}\sin(q_\perp x)e^{-t/\tau_{\rm s}}$, which describes the director $\mathbf{n} = (\sin(\varphi_{\rm s}), 0, \cos(\varphi_{\rm s}))$. These fluctuations are through the splay flexoelectric effect coupled with $\delta P$. However, deep in the $N_F$ phase, this coupling is small compared to the electrostatic term and effectively rescales $K_1$, so, for simplicity, we will neglect it. If $\rho_{\rm f} = 0$, the relevant part of the free energy density describing pure splay fluctuations, up to quadratic terms in $\varphi_{\rm s}$, becomes

$$f = \frac{1}{2}K_1\left(\frac{\partial\varphi_{\rm s}}{\partial x}\right)^2 + \frac{P_0^2\varphi_{\rm s}^2}{2\varepsilon_\perp\varepsilon_0}. \tag{F6}$$

This yields the dynamic equation for $\varphi_{\rm s}$,

$$\eta_1\frac{\partial\varphi_{\rm s}}{\partial t} = K_1\frac{\partial^2\varphi_{\rm s}}{\partial x^2} + \frac{P_0^2\varphi_{\rm s}}{\varepsilon_\perp\varepsilon_0}, \tag{F7}$$

giving the splay relaxation rate

$$\frac{1}{\tau_{\rm s}} = \frac{K_1 q_\perp^2}{\eta_1} + \frac{P_0^2}{\eta_1\varepsilon_\perp\varepsilon_0}. \tag{F8}$$

Here $\eta_1$ is the splay viscosity [35]. The mean-square amplitude of the fluctuation modes can be calculated by using the equipartition theorem, yielding the amplitudes of the splay fluctuations

$$\langle\varphi_{0\rm s}^2\rangle = \frac{k_{\rm B}T}{V\left(K_1 q_\perp^2 + \frac{P_0^2}{\varepsilon_\perp\varepsilon_0}\right)}. \tag{F9}$$

The electrostatic self-interaction therefore causes the splay fluctuations to become faster and have smaller amplitudes than in the $N$ phase.

If free ions are present in the system, they partially screen the bound charges and reduce the electrostatic effects. Maximal screening occurs if it can be assumed that free ions are much faster than the director dynamics, so that their density is given by the Boltzmann distribution $\rho^{\pm} = \rho_0 e^{\mp e\Phi/k_B T}$, then $\Phi$ can be calculated by the linearized Poisson-Boltzmann equation $\nabla \cdot \boldsymbol{\varepsilon}\nabla\Phi - \kappa^2\Phi = \nabla \cdot \mathbf{P}/\varepsilon_0$, where $\kappa^2 = \frac{2\rho_0 e}{\varepsilon_0 k_B T}$, $\rho_0$ is the density of free ions, and $e$ their charge. The splay relaxation rate and average square amplitude in this case become

$$\frac{1}{\tau_{s,i}} = \frac{K_1 q_\perp^2}{\eta_1} + \frac{P_0^2 q_\perp^4}{\eta_1 \varepsilon_\perp \varepsilon_0 (q_\perp^2 + \kappa_\perp^2)^2},$$
$$\langle \varphi_{0s,i}^2 \rangle = \frac{k_B T}{V\left(K_1 q_\perp^2 + \frac{P_0^2 q_\perp^4}{\varepsilon_\perp \varepsilon_0 (q_\perp^2 + \kappa_\perp^2)^2}\right)}. \quad \text{(F10)}$$

Here, $\kappa_\perp^2 = \kappa^2/\varepsilon_\perp$.

When free ions are not much faster than the director, then their dynamics can be described by the Nernst-Planck equation, which, in the absence of material flow, can be written as $\frac{\partial \rho^{\pm}}{\partial t} = \nabla \cdot (D\nabla\rho^{\pm} \pm \frac{De}{k_B T}\rho^{\pm}\nabla\Phi)$. Here, $D$ is the ion diffusion constant, which is assumed to be isotropic and identical for both anions and cations to obtain an analytical solution. These two equations, together with the dynamic equation for $\mathbf{n}$ and the Poisson equation $\nabla \cdot \boldsymbol{\varepsilon}\nabla\Phi = \nabla \cdot \mathbf{P}/\varepsilon_0 - (\rho^+ - \rho^-)/\varepsilon_0$, form a set of coupled equations. For small fluctuations in $\mathbf{n}$ and $\rho^{\pm}$ about the equilibrium structure with $\rho^{\pm} = \rho_0$, the coupled equations can be linearized. It can be shown that small periodic fluctuations of the angle $\varphi_s = \varphi_{0s}\sin(q_\perp x)e^{-t/\tau_s}$ are then coupled to small fluctuations of cation and anion densities $\delta\rho_n^{\pm} = \pm\rho_a \cos(q_\perp x)e^{-t/\tau_s}$, with the relaxation rates

$$\frac{1}{\tau_{s1,s2}} = \frac{K_1 q_\perp^2}{2\eta_1} + \frac{D(q_\perp^2 + \kappa^2)}{2} + \frac{P_0^2}{2\eta_1\varepsilon_\perp\varepsilon_0} \pm \frac{P_0^2}{2\eta_1\varepsilon_\perp\varepsilon_0}\sqrt{1 + \frac{4\varepsilon_\perp\varepsilon_0 q_\perp^2 (K_1 - D\eta_1)}{P_0^2} + \frac{4\varepsilon_\perp^2\varepsilon_0^2\left[K_1 q_\perp^2 - D\eta_1(q_\perp^2 + \kappa^2)\right]^2}{P_0^4}}. \quad \text{(F11)}$$

The relaxation rates of the pure twist fluctuations, i.e., small periodic fluctuations of the angle $\varphi_t = \varphi_{0t}\sin(q_\perp x)e^{-t/\tau_t}$, describing the director $\mathbf{n} = (0, \sin(\varphi_t), \cos(\varphi_t))$ are the same as in the $N$ phase:

$$\frac{1}{\tau_t} = \frac{K_2 q_\perp^2}{\gamma_1}, \quad \text{(F12)}$$

and the average square amplitudes are

$$\langle \varphi_{0t}^2 \rangle = \frac{k_B T}{V K_2 q_\perp^2}. \quad \text{(F13)}$$

In DLS experiments, the scattered intensity of a given mode is proportional to the average square amplitude of the mode and the so-called geometric factor, which depends on the scattering geometry [35]. In a typical scattering geometry for the measurement of the pure splay mode in our setup, $q_\perp = 6$–$8\ \mu\text{m}^{-1}$ [37]. In this geometry, in addition to the splay mode, the twist mode also contributes; however, when the splay and twist modes have comparable amplitudes, as is the case in the $N$ phase, the scattered intensity of the twist mode is due to the geometric factors [37] much smaller (typically a few percent of the splay mode intensity), and can be neglected. In the $N_F$ phase in this geometry, the scattered intensity was very low, and the measured relaxation rates corresponded to the twist mode. No other mode was distinguishable. In the absence of ions, the splay mode [Eq. (F8)] is, due to the electrostatic contribution alone, more than 4 orders of magnitude faster than the twist mode [Eq. (F12)], and its average square amplitude [Eq. (F9)] is more than 4 orders of magnitude smaller than that of the twist [Eq. (F13)]. In the sample F7i used in the measurements, the ion concentration was about $8 \times 10^{23}/\text{m}^3$, and if we assume maximal screening [Eq. (F10)], this amount of ions should be sufficient to suppress the electrostatic effects. However, the ions are too slow to efficiently screen the field. The ion diffusion constant is typically smaller than that calculated from the Stokes-Einstein equation [35], $D < k_B T/(6\pi\eta R_g) \approx 4 \times 10^{-12}\ \text{m}^2/\text{s}$ (here the ion radius of gyration $R_g \approx 0.5$ nm and shear viscosity $\eta \approx 0.1$ Pa s were taken), and is much smaller than $K_1/\eta_1 \gtrsim 10^{-10}\ \text{m}^2/\text{s}$ (assuming $K_1 \gtrsim 10$ pN and $\eta_1 \approx 0.1$ Pa s). Furthermore, $P_0^2/(\varepsilon_\perp\varepsilon_0) \gg K_1 q_\perp^2$ and $P_0^2/(\eta_1\varepsilon_\perp\varepsilon_0) \gg D(q_\perp^2 + \kappa^2)$, and consequently, the second and third terms under the square root in Eq. (F11) are much smaller than 1, so the relaxation rate of the director splay mode $1/\tau_{s1} \approx \frac{P_0^2}{\varepsilon_\perp\varepsilon_0\eta_1} \approx 50$ MHz is practically the same as in the ion-free case [Eq. (F8)], while $1/\tau_{s2} \approx D(q_\perp^2 + \kappa^2/2)$ is the relaxation rate of the ion fluctuation mode coupled to it. In addition to the coupled mode, the ions also exhibit an uncoupled fluctuating mode with $1/\tau_u = Dq_\perp^2$, in which the positive and negative ion concentrations fluctuate in phase. In DLS and DDM experiments, only director modes can be observed but, because in the $N_F$ phase the splay fluctuation modes are too fast and have very small amplitudes, they are not observed. In BDS experiments, we expect both modes in Eq. (F11) should be observed in the limit $q = 0$: the first one is the splay mode $m_{\text{NF,L}}$ discussed in the main text, which is a reorientational (Goldstone) mode, while the second one is the ion mode coupled to the splay mode. However, the finite sample thickness and surface properties modify the electrostatics of the system in this limit and, consequently, the relaxation rates measured in BDS [31]. Moreover, since the sample used for the dielectric measurements was not ion doped, $\kappa$ is expected to be small.

[1] R. J. Mandle, S. J. Cowling, and J. W. Goodby, A nematic to nematic transformation exhibited by a rod-like liquid crystal, Phys. Chem. Chem. Phys. **19**, 11429 (2017).

[2] H. Nishikawa, K. Shiroshita, H. Higuchi, Y. Okumura, Y. Haseba, S.-i. Yamamoto, K. Sago, and H. Kikuchi, A fluid liquid-crystal material with highly polar order, Adv. Mater. **29**, 1702354 (2017).

[3] N. Sebastián, L. Cmok, R. J. Mandle, M. R. de la Fuente, I. D. Olenik, M. Čopič, and A. Mertelj, Ferroelectric-ferroelastic phase transition in a nematic liquid crystal, Phys. Rev. Lett. **124**, 037801 (2020).

[4] X. Chen, E. Korblova, D. Dong, X. Wei, R. Shao, L. Radzihovsky, M. A. Glaser, J. E. Maclennan, D. Bedrov, D. M. Walba, and N. A. Clark, First-principles experimental demonstration of ferroelectricity in a thermotropic nematic liquid crystal: Polar domains and striking electro-optics, Proc. Natl. Acad. Sci. USA **117**, 14021 (2020).

[5] N. Sebastián, M. Čopič, and A. Mertelj, Ferroelectric nematic liquid-crystalline phases, Phys. Rev. E **106**, 021001 (2022).

[6] P. Kumari, B. Basnet, M. O. Lavrentovich, and O. D. Lavrentovich, Chiral ground states of ferroelectric liquid crystals, Science **383**, 1364 (2024).

[7] Y. Song, S. Aya, and M. Huang, Updated view of new liquid-matter ferroelectrics with nematic and smectic orders, Giant **19**, 100318 (2024).

[8] H. Kikuchi, H. Matsukizono, K. Iwamatsu, S. Endo, S. Anan, and Y. Okumura, Fluid layered ferroelectrics with global $C_{\infty v}$ symmetry, Adv. Sci. **9**, 2202048 (2022).

[9] J. Hobbs, C. J. Gibb, and R. J. Mandle, Emergent antiferroelectric ordering and the coupling of liquid crystalline and polar order, Small Sci. **4**, 2400189 (2024).

[10] J. Hobbs, C. J. Gibb, D. Pociecha, J. Szydłowska, E. Górecka, and R. J. Mandle, Polar order in a fluid like ferroelectric with a tilted lamellar structure–Observation of a polar smectic C ($SmC_P$) phase, Angew. Chem. Int. Ed. **64**, e202416545 (2024).

[11] D. Pociecha, J. Szydlowska, N. Vaupotič, K. Kwiatkowska, M. Juodka, J. Spiess, J. M. Storey, C. T. Imrie, R. Walker, and E. Gorecka, Twist grain boundary phases in proper ferroelectric liquid crystals realm, Adv. Sci. **12**, e08405 (2025).

[12] J. Karcz, J. Herman, N. Rychłowicz, P. Kula, E. Górecka, J. Szydlowska, P. W. Majewski, and D. Pociecha, Spontaneous chiral symmetry breaking in polar fluid–heliconical ferroelectric nematic phase, Science **384**, 1096 (2024).

[13] H. Nishikawa, D. Okada, D. Kwaria, A. Nihonyanagi, M. Kuwayama, M. Hoshino, and F. Araoka, Emergent ferroelectric nematic and heliconical ferroelectric nematic states in an achiral “straight” polar rod mesogen, Adv. Sci. **11**, 2405718 (2024).

[14] C. J. Gibb, J. Hobbs, D. I. Nikolova, T. Raistrick, S. R. Berrow, A. Mertelj, N. Osterman, N. Sebastián, H. F. Gleeson, and R. J. Mandle, Spontaneous symmetry breaking in polar fluids, Nat. Commun. **15**, 5845 (2024).

[15] J. Hobbs, C. J. Gibb, and R. J. Mandle, Ferri- and ferro-electric switching in spontaneously chiral polar liquid crystals, Nat. Commun. **16**, 7510 (2025).

[16] M. Cestari, S. Diez-Berart, D. A. Dunmur, A. Ferrarini, M. R. de la Fuente, D. J. B. Jackson, D. O. Lopez, G. R. Luckhurst, M. A. Perez-Jubindo, R. M. Richardson, J. Salud, B. A. Timimi, and H. Zimmermann, Phase behavior and properties of the liquid-crystal dimer 1″, 7″-bis(4-cyanobiphenyl-4′-yl) heptane: A twist-bend nematic liquid crystal, Phys. Rev. E **84**, 031704 (2011).

[17] K. Adlem, M. Čopič, G. R. Luckhurst, A. Mertelj, O. Parri, R. M. Richardson, B. D. Snow, B. A. Timimi, R. P. Tuffin, and D. Wilkes, Chemically induced twist-bend nematic liquid crystals, liquid crystal dimers, and negative elastic constants, Phys. Rev. E **88**, 022503 (2013).

[18] V. Borshch, Y.-K. Kim, J. Xiang, M. Gao, A. Jákli, V. P. Panov, J. K. Vij, C. T. Imrie, M. G. Tamba, G. H. Mehl, and O. D. Lavrentovich, Nematic twist-bend phase with nanoscale modulation of molecular orientation, Nat. Commun. **4**, 2635 (2013).

[19] A. G. Vanakaras and D. J. Photinos, A molecular theory of nematic–nematic phase transitions in mesogenic dimers, Soft Matter **12**, 2208 (2016).

[20] A. G. Vanakaras, E. T. Samulski, and D. J. Photinos, Polar-twisted, nano-modulated nematics: Form chirality and physical properties, Liquids **4**, 768 (2024).

[21] S. M. Shamid, S. Dhakal, and J. V. Selinger, Statistical mechanics of bend flexoelectricity and the twist-bend phase in bent-core liquid crystals, Phys. Rev. E **87**, 052503 (2013).

[22] P. De Gregorio, E. Frezza, C. Greco, and A. Ferrarini, Density functional theory of nematic elasticity: Softening from the polar order, Soft Matter **12**, 5188 (2016).

[23] L. Longa and G. Pająk, Modulated nematic structures induced by chirality and steric polarization, Phys. Rev. E **93**, 040701(R) (2016).

[24] L. Longa and W. Tomczyk, Twist–bend nematic phase from the Landau–de Gennes perspective, J. Phys. Chem. C **124**, 22761 (2020).

[25] M. Čopič and A. Mertelj, Q-tensor model of twist-bend and splay nematic phases, Phys. Rev. E **101**, 022704 (2020).

[26] J. V. Selinger, Director deformations, geometric frustration, and modulated phases in liquid crystals, Annu. Rev. Condens. Matter Phys. **13**, 49 (2022).

[27] P. M. Rupnik, E. Hanžel, M. Lovšin, N. Osterman, C. J. Gibb, R. J. Mandle, N. Sebastián, and A. Mertelj, Antiferroelectric order in nematic liquids: Flexoelectricity versus electrostatics, Adv. Sci. **12**, 2414818 (2025).

[28] X. Chen, V. Martinez, E. Korblova, G. Freychet, M. Zhernenkov, M. A. Glaser, C. Wang, C. Zhu, L. Radzihovsky, J. E. Maclennan, D. M. Walba, and N. A. Clark, The smectic $Z_A$ phase: Antiferroelectric smectic order as a prelude to the ferroelectric nematic, Proc. Natl. Acad. Sci. USA **120**, e2217150120 (2023).

[29] A. Sterle, P. M. Rupnik, L. Cmok, A. Erkoreka, M. Lavrič, N. Osterman, C. J. Gibb, J. Hobbs, J. Martinez-Perdiguero, R. J. Mandle, A. Mertelj, and N. Sebastián, Elasticity-driven periodic polarization patterns in confined chiral ferroelectric nematic fluid, Adv. Mater. e75002 (2026).

[30] E. E. Zavvou, A. Jarosik, H. Nádasi, R. Tuffin, P. K. Karahaliou, C. Krontiras, M. Klasen-Memmer, and A. Eremin, Signatures of polar order in a ferroelectric nematic liquid crystal: Splay stiffening and twist softening, Soft Matter **21**, 7508 (2025).

[31] A. Erkoreka, J. Martinez-Perdiguero, R. J. Mandle, A. Mertelj, and N. Sebastián, Dielectric spectroscopy of a ferroelectric nematic liquid crystal and the effect of the sample thickness, J. Mol. Liq. **387**, 122566 (2023).

[32] A. Erkoreka, A. Mertelj, M. Huang, S. Aya, N. Sebastián, and J. Martinez-Perdiguero, Collective and non-collective molecular dynamics in a ferroelectric nematic liquid crystal studied by

broadband dielectric spectroscopy, J. Chem. Phys. **159**, 184502 (2023).

[33] N. A. Clark, X. Chen, J. E. MacLennan, and M. A. Glaser, Dielectric spectroscopy of ferroelectric nematic liquid crystals: Measuring the capacitance of insulating interfacial layers, Phys. Rev. Res. **6**, 013195 (2024).

[34] *Broadband Dielectric Spectroscopy,* edited by F. Kremer and A. Schönhals (Springer, Berlin, Heidelberg, 2003).

[35] P.-G. de Gennes and J. Prost, *The Physics of Liquid Crystals*, 2nd ed., International Series of Monographs on Physics (Clarendon, Oxford, England, 1995).

[36] R. B. Meyer, Piezoelectric effects in liquid crystals, Phys. Rev. Lett. **22**, 918 (1969).

[37] N. Sebastián, L. Cmok, A. Petelin, R. J. Mandle, and A. Mertelj, Distinctive features of pretransitional behaviour between nematic phases as revealed by DDM, Liq. Cryst. **51**, 1047 (2024).

[38] I. Dozov, On the spontaneous symmetry breaking in the mesophases of achiral banana-shaped molecules, Europhys. Lett. **56**, 247 (2001).

[39] *Flexoelectricity in Liquid Crystals: Theory, Experiments and Applications,* edited by A. Buka and N. Eber (Imperial College Press, London, 2012).

[40] H. Nishikawa, D. Okada, D. Kwaria, A. Nihonyanagi, M. Kuwayama, M. Hoshino, and F. Araoka, CCDC 2356894: Experimental crystal structure determination (2025), doi: 10.5517/ccdc.csd.cc2k3jv2.

[41] H. Nishikawa, D. Kwaria, A. Nihonyanagi, and F. Araoka, 3D hierarchical twists in polar fluids: Chirality regulation by ultralow electric field, Adv. Mater. **38**, e13451 (2025).

[42] A. Erkoreka *et al.*, Supporting Data for Flexoelectricity-driven softening of bend elasticity leads to spontaneous chiral symmetry breaking in a polar fluid [Dataset], Zenodo, 2026, 10.5281/zenodo.20282283.

[43] A. Petelin, Cross-differential dynamic microscopy (Version V0.2.0) [Computer software], Zenodo, 2020, https://doi.org/10.5281/zenodo.3800382.

[44] J. C. Everts and M. Ravnik, Ionically charged topological defects in nematic fluids, Phys. Rev. X **11**, 011054 (2021).